\documentstyle[12pt,epsfig]{article}

\input{epsf}

\begin{document}

\pagestyle{empty}

\begin{flushright}
{\tt hep-ph/9706459}\\
{DFTT 37/97}\\
{Edinburgh 97/7}\\
{UB-ECM-PF 97/09}\\
\end{flushright}
\vskip 12pt
\begin{center}
\title{}

{\bf ASYMPTOTIC FREEDOM AT SMALL $x$} 
\vskip 24pt
{Richard D. Ball\footnote[1]{Royal Society University Research Fellow}}\\
\vskip 12pt
{\em Department of Physics and Astronomy}\\
{\em University of Edinburgh, EH9 3JZ, Scotland}\\
\vskip 18pt
{and}
\vskip 18pt
{Stefano Forte\footnote[2]{IBERDROLA visiting professor}}\\
\vskip 12pt
{\em INFN, Sezione di Torino,}\\
{\em Via P. Giuria 1, I-10125 Torino, Italy}\\
{\em and}\\
{\em Departament ECM, Universitat de Barcelona,}\\
{\em Diagonal 647, E-08028 Barcelona, Spain}\\ 

\vskip 36pt
\end{center}
{\begin{center}{\bf Abstract}\\
\noindent We describe how perturbative QCD may be applied to inelastic $e$-$p$ 
scattering at high center-of-mass energies, 
i.e. at small $x$ and fixed $Q^2$.\end{center}  
}
\vskip 36pt
\begin{center}
{Talk at {\em DIS97}, Chicago, April 1997}\\
\smallskip
{\em to be published in the proceedings}\\
\end{center}
\bigskip
\begin{flushleft}
{June 1997}
\end{flushleft}
\eject

\pagenumbering{arabic}
\pagestyle{plain}

\def\toinf#1{\mathrel{\mathop{\sim}\limits_{\scriptscriptstyle
{#1\rightarrow\infty }}}}
\def\tozero#1{\mathrel{\mathop{\sim}\limits_{\scriptscriptstyle
{#1\rightarrow0 }}}}
\def\frac#1#2{{{#1}\over {#2}}}
\def\half{\hbox{${1\over 2}$}}\def\third{\hbox{${1\over 3}$}}
\def\quarter{\hbox{${1\over 4}$}}
\def\smallfrac#1#2{\hbox{${{#1}\over {#2}}$}}
\def\as{\alpha_s}
\def\bas{\bar\alpha_s}
\def\asmu{\alpha_s(\mu^2)}
\def\asQ{\alpha_s(Q^2)}
\def\asS{\alpha_s(S^2)}
\def\Lam{\Lambda}
\def\muderiv{\mu^2\frac{\partial}{\partial\mu^2}}
\def\Qderiv{Q^2\frac{\partial}{\partial Q^2}}
\def\Sderiv{S^2\frac{\partial}{\partial S^2}}
\def\rgederiv{\muderiv+\beta(\asmu)\frac{\partial}{\partial\alpha_s}}
\def\PR{{\it Phys.~Rev.~}}
\def\PRL{{\it Phys.~Rev.~Lett.~}}
\def\NP{{\it Nucl.~Phys.~}}
\def\NPBPS{{\it Nucl.~Phys.~B (Proc.~Suppl.)~}}
\def\PL{{\it Phys.~Lett.~}}
\def\PRep{{\it Phys.~Rep.~}}
\def\AP{{\it Ann.~Phys.~}}
\def\CMP{{\it Comm.~Math.~Phys.~}}
\def\JMP{{\it Jour.~Math.~Phys.~}}
\def\NC{{\it Nuov.~Cim.~}}
\def\SJNP{{\it Sov.~Jour.~Nucl.~Phys.~}}
\def\SPJETP{{\it Sov.~Phys.~J.E.T.P.~}}
\def\ZP{{\it Zeit.~Phys.~}}
\def\JP{{\it Jour.~Phys.~}}
\def\vol#1{{\bf #1}}\def\vyp#1#2#3{\vol{#1} (#2) #3}
\catcode`@=11 %This allows us to modify plain macros
\renewcommand\section{\@startsection {section}{1}{\z@}
    {-3.5ex plus -1ex minus -.2ex}{2.3ex plus .2ex}{\bf}}
\renewcommand\subsection{\@startsection {subsection}{1}{\z@}
    {-3.5ex plus -1ex minus -.2ex}{2.3ex plus .2ex}{\it}}
\def\slash#1{\mathord{\mathpalette\c@ncel#1}}
 \def\c@ncel#1#2{\ooalign{$\hfil#1\mkern1mu/\hfil$\crcr$#1#2$}}
\def\lsim{\mathrel{\mathpalette\@versim<}}
\def\gsim{\mathrel{\mathpalette\@versim>}}
 \def\@versim#1#2{\lower0.2ex\vbox{\baselineskip\z@skip\lineskip\z@skip
       \lineskiplimit\z@\ialign{$\m@th#1\hfil##$\crcr#2\crcr\sim\crcr}}}
\catcode`@=12 %at signs are no longer letters
\newcommand{\ccaption}[2]{\begin{center}\parbox{0.85\textwidth}
           {\caption[#1]{\small{\it{#2}}}}\end{center}}

The structure functions for inclusive inelastic 
$e$-$p$ scattering depend on two scales: the virtuality $Q^2$ of the photon
and the center-of-mass energy $W^2$ of the photon-proton collision.
The Bjorken variable is defined as essentially the ratio of these two: 
$x=Q^2/(W^2+Q^2-m_p^2)$. If neither scale is large, it is generally 
agreed that a pertuyrbative computation of the cross-section is not possible.
If $Q^2$ becomes very large with $x$ fixed, mass factorization (or the OPE)
and renormalization group invariance together imply that the appropriate scale 
for the running of the coupling is $Q^2$, and asymptotic freedom then 
justifies perturbative evolution in $Q^2$ of moments with respect
to $x$ (see fig.~1).

At large $Q^2$ structure functions \cite{bdr} display Bjorken scaling, 
i.e. they are essentially flat, with logarithmic scaling violations. 
Structure functions are also essentially flat at fixed $Q^2$ 
but large $W^2$  
(and thus small $x$), but as yet no entirely satisfactory explanation
exists. The gentle growth of all physical cross-sections appears to be 
universal, and is often parameterized as $x^{\alpha-1}$, where $\alpha$ 
is the intercept of the `pomeron' singularity, and consequently close to 
unity. Logarithmic parameterizations are also acceptable, however. Finally, 
when both $Q^2$ and $W^2$ are large, but $x$ is small, structure functions   
rise quite steeply with both $x$ and $Q^2$, in accordance with the 
double scaling prediction of perturbative QCD \cite{DGPTWZ}, and 
consistent with the flat behaviour on either side.
% (see fig.~1). 

The previous paragraph summarises the conventional theory of structure 
functions within perturbative QCD at high virtuality and Regge theory at 
high energy. To improve on this simple picture it is tempting to try to
somehow derive the pomeron directly from QCD. 
At small $x$ the usual perturbative expansion is dominated by terms of 
the form $\alpha_s^n\ln^n 1/x$, and if these are summed at fixed $Q^2$ and
fixed coupling they indeed generate a powerlike 
rise\cite{BFKL}. However the relatively large value of the intercept
$\alpha$ found in these calculations  not only disagrees with the 
much softer rise in the data at low $Q^2$, but also spoils double 
scaling \cite{DAS}. Furthermore, if the extra logarithms are included 
in the perturbative evolution to higher $Q^2$ \cite{CH},
they result in a strong factorization scheme dependence\cite{Sch}, 
and in general generate too steep a rise in $x$ and 
$Q^2$\cite{DSV}. It thus seems that 
at best the higher order logarithms provide a poor approximation
to the higher order terms in the conventional perturbation series: 
this may perhaps have been expected, since the series of logarithms 
converges, while the complete perturbative expansion is at best asymptotic. 

The breakdown of the conventional formulation of perturbative QCD
(based on the resummation of logarithms of $Q^2$) in the small $x$ limit
thus cannot be cured by summing the large logarithms of $1/x$.
However since the Regge limit involves a single large scale, 
$W^2\gg Q^2$, there remains 
the possibility that perturbative QCD might still work provided it is 
reformulated in such a way that the evolution is with respect to $W^2$ 
(or $x$) and applies to moments with respect to $Q^2$. 
This will require a new factorization theorem,
giving rise to novel evolution equations and parton distributions\cite{afp}.

In a physical gauge a two particle reducible contribution to a physical 
cross-section may be factorized \cite{CH} into an infrared finite 
coefficient function,
depending on the renormalization scale $\mu$ only through the renormalised 
coupling $\asmu$, and an `unintegrated' parton distribution function
$f(\frac{l^2}{\mu^2},\frac{k^2}{\mu^2};\mu^2)$ which gives the distribution 
of partons in the hadron according to their longitudinal and 
transverse momenta in the infinite momentum frame:
\begin{equation}
\sigma^{(2)}\Big(\frac{S^2}{\mu^2},\frac{Q^2}{\mu^2};\mu^2\Big)=
\int_0^\infty \frac{dl^2}{l^2}\int_0^\infty \frac{dk^2}{k^2}
C\Big(\frac{S^2}{l^2},\frac{Q^2}{k^2};\asmu\Big)
f\Big(\frac{l^2}{\mu^2},\frac{k^2}{\mu^2};\mu^2\Big),
\label{doubfac}
\end{equation}
where $S^2\equiv\Lam^2/x$, $l^2\equiv\Lam^2/y$, and $\Lam$ is some fixed
scale typical of the strong interaction. Since all the $\mu$
dependence is contained in the coupling, the separation of the 
integral over the momenta of the intermediate parton into 
longitudinal and transverse convolutions is entirely kinematic. The 
convolutions can be undone by Mellin transforms, to give
\begin{equation}
\sigma_{NM}^{(2)}=C_{NM}(\asmu)f_{NM}(\mu^2).
\end{equation}

At large $Q^2$, the double factorization eq.(\ref{doubfac}) may be used to 
derive the more familiar mass factorization 
\begin{equation}
\sigma_N(Q^2/\mu^2;\mu^2) = C_N(Q^2/\mu^2;\asmu) F_N(\mu^2) + O(1/Q^2),
\label{massfac}
\end{equation}
where now $F_N(\mu^2)$ is the Mellin transform of a parton 
distribution $F(y,\mu^2)$, which can be expressed in terms of the
logarithmic moments with respect to $k^2$ of the original unintegrated 
distribution. Multiparton (i.e. two particle irreducible) contributions 
to the cross-section are suppressed by powers of $Q^2$ (higher twist).
Renormalization group invariance of the right hand side of eq.(\ref{massfac})
then gives the usual evolution equation for the parton distribution and 
renormalization group equation for the coefficient function, whose solution 
leads to the choice of $Q^2$ as an appropriate factorization scale.

At large $S^2$ (i.e. small $x$), because the double factorization 
eq.(\ref{doubfac}) is essentially symmetrical in transverse and longitudinal 
scales, a similar argument gives us instead the energy factorization 
\cite{afp} 
\begin{equation}
\sigma_M(S^2/\mu^2;\mu^2) = C_M(S^2/\mu^2;\asmu) F_M(\mu^2)+
\sigma^q_M(S^2)+O(1/S^2).
\label{energyfac}
\end{equation}
Here $F_M(\mu^2)$ is the Mellin transform of a different integrated parton 
distribution function $F(k^2,\mu^2)$, which gives the distribution of 
partons in transverse momentum, and can be expressed in terms of 
logarithmic moments with respect to $l^2$ of the unintegrated distribution.
Multiparton contributions are however no longer suppressed by powers of $S^2$,
though they cannot grow with $S^2$ because collinear and soft singularities
only arise from emissions from a single parton \cite{EGMPR}. They may 
thus be ignored asymptotically, even though at practical energy scales 
they may constitute an important background to the leading single parton
contribution. 

Renormalization group invariance of the right hand side of 
eq.(\ref{energyfac}) now gives an evolution equation  
\begin{equation}
\muderiv F_M(\mu^2) = \gamma_M(\asmu)F_M(\mu^2),
\label{longev}
\end{equation}
for the integrated parton distribution $F_M(\mu^2)$, and 
a renormalization group equation for the coefficient function, 
which may be solved in the usual way to yield
\begin{equation}
\sigma_M(S^2/\mu^2;\mu^2) = C_M(1;\asS) F_M(S^2)
+ \sigma^q_M(S^2)+ O(1/S^2).
\label{longrge}
\end{equation}
The appropriate choice of factorization scale at small $x$ is thus $S^2$: 
QCD becomes asymptotically free as $x\to 0$, and perturbative calculations 
of the coefficient function and evolution of the parton distribution 
may be performed self-consistently. This is in contrast to the more usual
approach \cite{BFKL,CH} in which the logarithms of $x$ are added
to the usual evolution in $Q^2$: there the coupling is either fixed or 
runs with $Q^2$, and the calculation becomes inconsistent in the high 
energy limit since the coupling is no longer falling to compensate for the 
large logarithms.

The anomalous dimension $\gamma_M(\as)$ in the longitudinal evolution equation 
(\ref{longev}) may be thought of as the Mellin transform of a longitudinal 
splitting function. This may be calculated in a very similar way to the 
Altarelli--Parisi calculation of the usual (transverse) splitting 
function, by considering parton emission in the Weizs\"acker--Williams 
approximation, but now with strong ordering of longitudinal rather than
transverse momenta. The computation of LO graphs is then sufficient to
determine the cross section at the leading logarithmic level to all
orders in the coupling. Since large logarithms are now generated by 
both $s$ and $t$ channel emissions, it is necessary to calculate the 
leading log contribution from the LO
graphs in fig.~2a. The virtual graph cancels the infrared singularity 
in the real emissions, and the final result is \cite{afp}
\begin{equation}
\gamma_M(\as)=C_A\frac{\as}{\pi}[2\psi(1)-\psi(M)-\psi(1-M)]+O(\as^2).
\label{anomdim}
\end{equation}
It follows that if the coupling were fixed, at LO the longitudinal evolution 
equation (\ref{longev}) would reduce to the BFKL equation if 
in addition distinctions between integrated and unintegrated distributions 
were ignored.

Since quarks make no contribution to the LO splitting function, only the gluon
distribution actually evolves:
the quark contribution to the cross-section 
may thus be included in the second (background) term in eq.(\ref{longrge}).
Coefficients functions for $F_2$ and $F_L$ may be deduced from the quark
box calculations in ref.\cite{CH}. Only ratios of coefficient functions 
have physical significance, since an arbitrary factor may be 
absorbed into the definition of the gluon distribution.

In fig.~2 we also show some of the graphs which contribute to 
the emission amplitude at NLO. Real emissions fig.~2b contain  both  leading 
logarithms which are iterations of the LO real emissions in fig.~2a
(and are thus reproduced when solving the LO evolution equations), but 
also next-to-leading logarithms, i.e. contributions to the NLO
anomalous dimension. The one-loop corrections fig.~2c further  contain
leading logarithms which are not obtained by simple iteration of the LO graphs,
but are instead proportional to $\beta_0$ and are thus generated  by
the running of the coupling with $S^2$ in the LO evolution. This follows 
as a direct consequence of the Callan-Symanzik equation satisfied by
the gluon  four-point function; notice that the 
virtuality of the $s$-channel gluon in the first two graphs is
$k^2/y$. Finally the two-loop graphs fig.~2d are necessary to 
cancel infrared singularities. There is furthermore a set of graphs involving 
quark loops and production of $q$-$\bar{q}$ pairs. \footnote{It is not 
clear to us in precisely which circumstances all these graphs might be 
equivalent to the `Reggeized' graphs calculated in  
ref.\cite{Fadin,CC}. What is clear however is that a NLO calculation 
can only be useful when corrections 
to both splitting functions and coefficient functions have been computed 
in the same renormalization and factorization scheme.} 
%Use of the renormalization group however
%organizes this cumbersome set of diagrams in a simple and compact way.

We may now consider some of the implications of the longitudinal evolution
eq.(\ref{longev}). At high energies we will find logarithmic 
scaling violations, 
just as with the usual transverse evolution at high virtualities. But 
because the longitudinal anomalous anomalous dimension has a minimum at 
$M=\half$, rather than decreasing monotonically, and because there is 
only one evolving parton, all (singlet) structure functions will exhibit 
the same universal behaviour at high energies:
\begin{equation}
F(x,Q^2)\tozero{x}{\cal N} \left(\frac{Q^2}{\Lam_0^2}\right)^{\half}
\left(\ln\ln{ 1\over x}\right)^{-\half}
\left(\ln{ 1\over x}\right)^{4\ln 2\gamma^2-1}.
\end{equation}
In this way perturbative QCD reproduces the gentle growth in the high energy 
cross-section, with no need for a Regge trajectory with intercept greater 
than unity. Being logarithmic, the growth is sufficiently gentle to 
match smoothly onto the double scaling behaviour at higher $Q^2$, and 
all the difficulties found in \cite{bdr,Sch,DSV} are resolved. Furthermore
a careful examination of the $Q^2\to 0$ limit shows that 
the photoproduction cross-section remains finite, and there
is no violation of unitarity bounds. However in this limit 
our perturbative approach
may break down due to infrared logarithms in the same way that 
the usual analysis at large $Q^2$ breaks down as $x\to 1$ 
due to Sudakov logarithms. 

We are thus led naturally into a world without a pomeron. There,
we can approach a variety of problems from the vantage point of
perturbative QCD: besides the calculation of NLO corrections to 
structure functions, we 
might consider perturbative computations of
diffractive processes,  of high energy photon-photon 
cross-sections, and of high energy hadronic processes in which there is 
no large transverse scale. There remains much to be done. 

\section*{Acknowledgments}
 We would like to thank 
I.~Balitskii, V.~Del~Duca, V.S.~Fadin, E.~Levin, G.~Marchesini 
and G.~Sterman for various discussions and comments
during the workshop.

\begin{figure}[htbp] \centering        
\epsfig{file=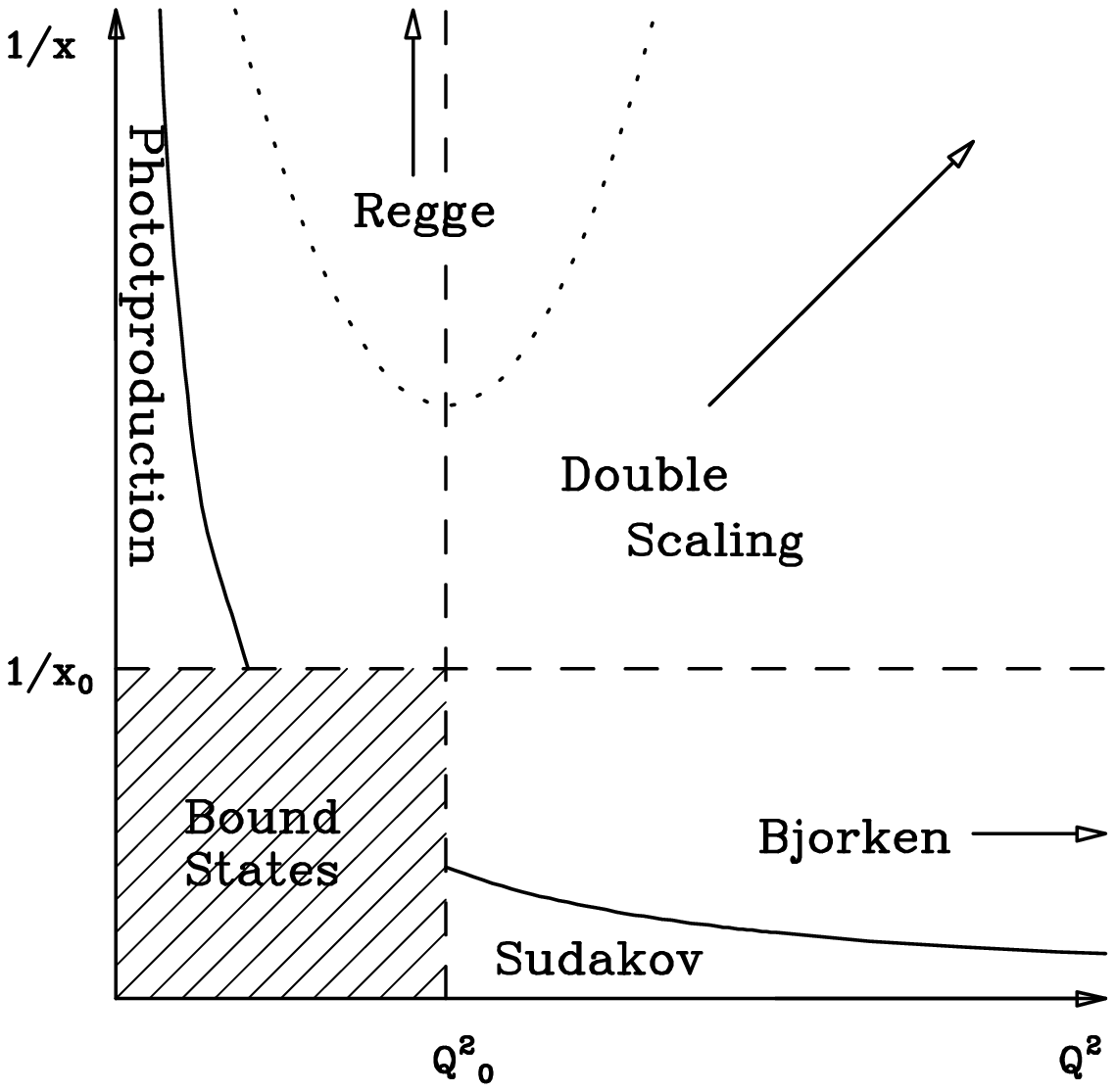,width=100mm}
\ccaption{}{Regions and limits in the $(Q^2,{1\over x})$ plane.
The shaded area denotes the region of nonperturbative dynamics where
there is no large scale. To the right of an initial condition set at 
$Q^2=Q_0^2$ we may evolve using the transverse evolution equation, 
while above an initial condition set at 
$x=x_0$ we may evolve using the longitudinal evolution equation.}
\label{fig1}
\vskip-0.2in
\end{figure}                      

\begin{figure}[hbp!] \vskip-0.2in\centering        
\epsfig{file=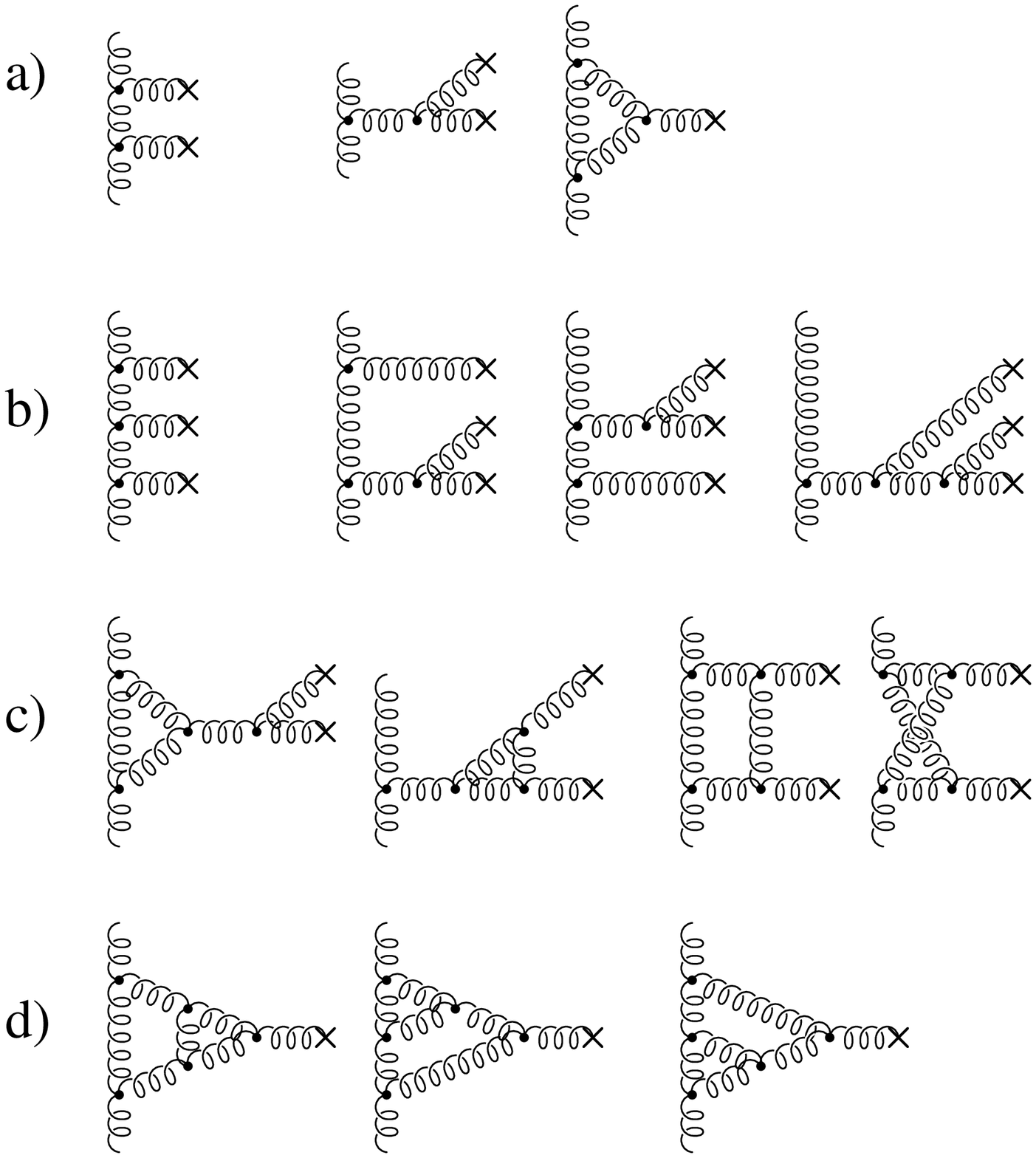,width=110mm}
\ccaption{}{Contributions to gluon emission amplitudes: a) LO contributions
b)-d) NLO contributions described in the text. There are also many 
contributions due to self energy insertions, and NLO contributions from 
emission of $q\bar{q}$ pairs and quark loops.}
\label{fig2}
\vskip0.0in
\end{figure}                      
 
\end{document}